\documentstyle[12pt,worldsci]{article}
\begin{document}
\def\z2{\ifmmode Z_2\else $Z_2$\fi}
\def\epem{\ifmmode e^+e^-\else $e^+e^-$\fi}
\def\gev{\,{\rm GeV}}
\def\ie{{\it i.e.},}
\def\eg{{\it e.g.},}
\def\etal{{\it et. al.}}
\def\to{\rightarrow}
\def\Re{{\cal R \mskip-4mu \lower.1ex \hbox{\it e}\,}}
\def\Im{{\cal I \mskip-5mu \lower.1ex \hbox{\it m}\,}}
\newskip\zatskip \zatskip=0pt plus0pt minus0pt
\def\matth{\mathsurround=0pt}
\def\lsim{\mathrel{\mathpalette\atversim<}}
\def\gsim{\mathrel{\mathpalette\atversim>}}
\def\atversim#1#2{\lower0.7ex\vbox{\baselineskip\zatskip\lineskip\zatskip
  \lineskiplimit 0pt\ialign{$\matth#1\hfil##\hfil$\crcr#2\crcr\sim\crcr}}}
\def\undertext#1{$\underline{\smash{\vphantom{y}\hbox{#1}}}$}
\pagestyle{empty}
\setlength{\baselineskip}{2.6ex}
\font\elevenbf=cmbx10 scaled\magstep 1

\title{{\bf VECTOR LEPTOQUARK PRODUCTION AT HADRON COLLIDERS}}
\vspace{0.4cm}
\author{J.L.~HEWETT, T.G.~RIZZO\\
{\em High Energy Physics Division, Argonne National Laboratory, Argonne, IL
60439}\\
\vspace{0.2cm}
S.~PAKVASA\\
{\em Department of Physics and Astronomy, University of Hawaii at Manoa,
Honolulu, HI 96822}\\
\vspace{0.2cm}
H.E.\ HABER, A.~POMAROL\\
{\em Santa Cruz Institute for Particle Physics, University of California,
Santa Cruz, CA  95064}
}
\maketitle

\begin{center}
\parbox{13.0cm}
{\begin{center} ABSTRACT \end{center}
{\small\hspace*{0.3cm}

We explore the production of vector leptoquarks($V$) at the Tevatron, LHC,
and SSC through both quark-antiquark and gluon fusion: $q \bar q, gg \to VV$.
The cross sections are found to be somewhat larger than for scalar leptoquarks
of the same mass implying enhanced search capabilities.}}

\end{center}

Many extensions of the standard model (SM) which place quarks and leptons on
an equal footing predict the existence of leptoquarks, which are spin-0 or 1
objects that couple to a $q\ell$ or $\bar q\ell$ pair{\cite {bigref}}. While
these objects may be sought indirectly through their influence on low energy
processes{\cite {sacha}}, the most promising approach is via direct production
at colliders. In particular, searches for leptoquarks at
LEP{\cite {lep}}, HERA{\cite {hera}}, and the Tevatron{\cite {tev}}
have already been performed, in most cases concentrating on the specific
scenario of scalar leptoquarks.  While the current LEP bounds ($M_{LQ}\gsim
M_Z/2$) are insensitive to the leptoquark spin, the HERA limits
display a sizeable sensitivity to the choice of spin-0 or spin-1.  The
HERA bounds also depend directly on the unknown $q\ell$-leptoquark coupling;
taking this coupling to be electroweak strength, the leptoquark mass is
restricted to be $M_{LQ}> 98-192\gev$, for various choices of the spin,
electric charge, and helicity of the leptoquark. The Tevatron places a
bound of $M_{LQ}>116\gev$ (assuming a $100\%$ branching ratio into electrons)
on scalar leptoquarks.  The sensitivity of the Tevatron limits to the spin
of the leptoquark has not yet been addressed in the literature.
The cross section for the $q \bar q \to VV$ subprocess, with $V$ denoting
a vector leptoquark, is readily obtainable from existing results, while the
parton level cross section for $gg \to VV$ remains to be calculated.
In this work we calculate these cross sections at the Tevatron, the LHC,
and the SSC,
and compare our results to those previously obtained for scalar leptoquarks
{\cite {scalar}}. We will show that the cross section in the spin-1 case may
be substantially larger than that for spin-0, implying stronger search limits
from existing data and an extended search range in the future.
We also find that while the $gg$ subprocess is generally dominant at both
the LHC and SSC, the $q \bar q$ subprocess at the Tevatron is extremely
important particularly for larger values of the $V$ mass. We note that
vector leptoquark production at high energy linear \epem\ colliders
has been considered elsewhere in the literature{\cite {vlq}}.

In order to calculate the $gg \to VV$ cross section we need to determine both
the trilinear $gVV$ and quartic $ggVV$ couplings, which may naively at first
appear to be unknown. (For the $q \bar q$ subprocess, only the $gVV$
coupling is required.) However, in any realistic model wherein vector
leptoquarks appear and are fundamental objects, they will be
the gauge bosons of an extended gauge group. In this case the $gVV$ and
$ggVV$ couplings are completely fixed by gauge invariance.  These
particular couplings will also insure that the subprocess cross section obeys
tree-level unitarity, as is the hallmark of all gauge theories.
Of course, it might be that the appearance of vector
leptoquarks is simply some low energy manifestation of a
more fundamental theory at a higher scale and that these particles may even
be composite, in which case so-called `anomalous' couplings in both the $gVV$
and $ggVV$ vertices can appear. One such possible coupling is an `anomalous
magnetic moment', usually described in the literature by the parameter $\kappa$
(Ref.\ 8), which takes the value of unity in the gauge theory case.
Among these `anomalous couplings', the term which induces $\kappa$ is
special in that it is the only one that conserves $CP$ and is of
dimension 4.

The Feynman rules for the leptoquark-gluon inteactions
are derived from the following effective Lagrangian which includes the
most general set of $CP$-conserving operators of dimension 4 (or less)
\begin{equation}
{\cal L}_V =-{1\over 2} F^\dagger_{\mu\nu}F^{\mu\nu}+M_V^2V^\dagger_\mu V^\mu
-ig_s\kappa V^\dagger_\mu G^{\mu\nu}V_\nu  \,.
\end{equation}
Here, $G_{\mu\nu}$ is the usual gluon field strength tensor,
$V_\mu$ is the leptoquark field and $F_{\mu\nu}=D_\mu V_\nu-D_\nu
V_\mu$, where $D_\mu=\partial_\mu+ig_sT^a G^a_\mu$ is the gauge
covariant derivative (with respect to $SU(3)$ color), $G^a_\mu$ is the
gluon field and the $SU(3)$ generator $T^a$ is taken in the triplet
representation.  As
values of $\kappa$ differing from one have been entertained in the literature
when discussing vector leptoquarks
{\cite {vlq}}, we will generally assume $\kappa=1$ or 0, with the latter
value corresponding to `minimal' coupling, in order to probe the sensitivity
of our results to the assumed gauge nature of $V$. We will also describe the
results in the more general case where $\kappa$ is arbitrary.

The calculation of the parton-level differential cross section,
$d\hat \sigma/d\hat t$,
for the $gg$ subprocess for arbitrary values of $\kappa$ is now
straightforward but algebraically cumbersome. The details of the calculation
will be presented elsewhere{\cite {huge}}.
The matrix element receives
contributions from $\hat s$-, $\hat t$-, and $\hat u$-channel graphs as well
as the $ggVV$ four-point seagull graph, as depicted in Fig.\ 1.  All of
these diagrams appear in the
corresponding scalar leptoquark case albeit with a different tensor structure.
Once the result for this parton level cross section is obtained
we can fold it together with the gluon distributions from the initial state
hadrons and impose either rapidity
and/or $p_t$ cuts. To this, the $q \bar q$ contribution must be added to
obtain the total cross section. The $q\bar q$ subprocess cross section can be
easily obtained from the process
$e^+e^- \to W^+W^-$ via a virtual photon, by the replacement
$\alpha \to \alpha_s$ and the incorporation of the appropriate
color factors. We remind the
reader that this cross section also depends on the choice of
$\kappa$. In addition to an $\hat s$-channel gluon exchange, shown in Fig.\ 2a,
this subprocess
can also receive contributions from $\hat t$- or $\hat u$-channel diagrams as
displayed in Fig.\ 2b.  These additional diagrams involve lepton exchange and
are proportional to the square of the unknown $q\ell$-leptoquark
Yukawa couplings.
If these couplings are assumed to be of electromagnetic strength or smaller,
we find that these diagrams do not make a significant contribution and are
hence neglected in our analysis.  (A similar situation, of course,
arises\cite{scalar} in the case of scalar leptoquarks.)

The two individual subprocess result in the total cross sections displayed
in Figs.\ 3, 4 and 5 at the Tevatron, SSC, and LHC, respectively, and should
be compared with the existing calculations for the scalar case in Ref.~6.
In obtaining these results we have made use of the NLO parton distributions
of Ref.~10. As we see from the
figures, the production rate for spin-1 leptoquarks can be substantially
larger than in
the spin-0 case. The current Tevatron limit of 116 GeV (Ref.\ 5)
on the mass of a hypothetical scalar leptoquark which  couples to only
the first generation fermions, can be
translated easily to the vector case if identical assumptions about the
leptoquark branching fractions are made.  We find, taking the
branching fraction of $V$ to the $\ell^{\pm}+j$ final state
to be 1(0.5), that the vector leptoquark mass must be $\gsim$ 230(200) GeV at
$95 \%$ CL assuming $\kappa=1$. A slightly smaller bound is obtained in
the case of $\kappa=0$, with $M_V\gsim 180(150)\gev$, showing that the
constraints on the mass of vector leptoquarks
from hadron colliders are somewhat sensitive to the assumption that
these particles are fundamental gauge bosons from an extended gauge model.
We anticipate that these limits will increase to
near 270(250) GeV for the $\kappa=1$ case after the data from the 1992-93
Tevatron run 1a have been fully analyzed. As we can see from
these calculations, the Tevatron may eventually be able to push the lower
bound on the
mass of $V$ beyond the kinematic reach of HERA, at least for the case of a
large $\ell^{\pm}+j$ branching fraction and $\kappa=1$.
These figures also show that at the Tevatron the $gg$ subprocess is
overcome by the $q \bar q$ subprocess for $V$ masses above 50 GeV. We also
learn that the $gg$ contribution has a minimum very close to
$\kappa=0$ and the corresponding minimum for
the $q \bar q$ subprocess is also not far from zero (but is somewhat
dependent upon the average $\hat s$ and the vector leptoquark mass). This
implies that the $\kappa=0$ result essentially yields
the smallest vector leptoquark pair cross section at the Tevatron
and can be used to place a model-independent bound.

At the SSC/LHC, the accessible mass range for vector leptoquarks is
also seen to be significantly
larger than in the scalar case. Since the contribution from the $q \bar q$
subprocess is insignificant here, the total cross section will have a minimum
near $\kappa=0$ so that this cross section can again be used to
set a model-independent limit on $V$ pair production.  Assuming
an integrated luminosity of $10 fb^{-1}$ at the SSC, and a 10 event
discovery limit, the mass reach for vector (scalar) leptoquarks is
3.6 (2.2) TeV assuming $\kappa=1$. A somewhat smaller search limit
of 3.0 TeV is obtained for $\kappa=0$, which again
demonstrates the sensitivity to the assumptions about the fundamental
nature of these particles. At the LHC (taking ${\sqrt s=14}$ TeV), we obtain
the corresponding search limits for vector leptoquarks of 2.2(1.8) for
$\kappa=1(0)$ assuming a factor of ten increase in the integrated
luminosity is available in comparison to the SSC. For scalar leptoquarks a
limit of 1.4 TeV is achievable at the LHC.

If leptoquark type events are discovered, it will be important
to determine the underlying nature of these events.
Observing a mass peak in the jet-lepton invariant mass would
rule out Standard Model backgrounds such as $W$+jet production.
It will then be crucial to measure the spin and color of the
leptoquark (to distinguish it from other possible exotic objects
such as a leptogluon).  This will clearly require a substantially
larger data sample than the 10 events referred to above.
In a future publication, we intend to study in detail the nature
of the leptoquark signals and ascertain the requirements for
a full determination of the leptoquark quantum numbers.  Clearly,
the discovery of such a particle would have a profound effect on
the development of theories beyond the Standard Model.

\vspace{0.6cm}
{\elevenbf\noindent Acknowledgements \hfil}
\vglue 0.4cm
This work has been supported in part by the U.S. Department of Energy,
Division of High Energy Physics. HEH acknowledges support from the Texas
National Research Laboratory Commission under Grant No. RGFY93-330.  JLH,
TGR, and SP would like to thank the LSGNA Collaboration for fruitful
discussions during this workshop.

\vspace{1.0cm}
%
\def\MPL #1 #2 #3 {Mod.~Phys.~Lett.~{\bf#1},\ #2 (#3)}
\def\NPB #1 #2 #3 {Nucl.~Phys.~{\bf#1},\ #2 (#3)}
\def\PLB #1 #2 #3 {Phys.~Lett.~{\bf#1},\ #2 (#3)}
\def\PR #1 #2 #3 {Phys.~Rep.~{\bf#1},\ #2 (#3)}
\def\PRD #1 #2 #3 {Phys.~Rev.~{\bf#1},\ #2 (#3)}
\def\PRL #1 #2 #3 {Phys.~Rev.~Lett.~{\bf#1},\ #2 (#3)}
\def\RMP #1 #2 #3 {Rev.~Mod.~Phys.~{\bf#1},\ #2 (#3)}
\def\ZP #1 #2 #3 {Z.~Phys.~{\bf#1},\ #2 (#3)}
\def\IJMP #1 #2 #3 {Int.~J.~Mod.~Phys.~{\bf#1},\ #2 (#3)}
\bibliographystyle{unsrt}

\newpage

{
\noindent
Fig.\ 1: The Feynman diagrams responsible for the parton level process
$gg\to VV$.

\medskip

\noindent
Fig.\ 2: Feynman diagrams for $VV$ production in $q\bar q$ scattering.

\medskip

\noindent
Fig.~3: Production cross section for a pair of vector leptoquarks at the
Tevatron as a function of the leptoquark mass assuming
(a)$\kappa=1$ or (b)$\kappa=0$.
The dotted(dashed, solid)curve corresponds to the $q \bar q$($gg$, total)
contribution respectively for the CTEQ1MS parton distributions. Our results
can vary by about 10 $\%$ if other distributions are used instead.
(c)$\kappa$ dependence of the $q \bar q$
(dots), $gg$(dashes), and total(solid) $V$ pair production cross sections
at the Tevatron for a vector leptoquark mass of 200 GeV assuming the CTEQ1MS
distributions. No cuts have been applied in this figure.

\medskip

\noindent
Fig.~4: Same as Fig.~3, but for the SSC. In (c), a vector leptoquark mass
of 1 TeV is assumed.

\medskip

\noindent
Fig.~5: Same as Fig.~4, but for the LHC.

 }

\end{document}